\documentclass[twocolumn,aps,prl,showpacs,floatfix]{revtex4}
\usepackage{graphicx}
\usepackage{subfigure}
\usepackage{epsfig}
\usepackage{psfrag}

\makeatletter

\usepackage{verbatim}

\makeatletter

\input epsf

\def\be{\begin{equation}}
\def\ee{\end{equation}}
\def\ba{\begin{eqnarray}}
\def\ea{\end{eqnarray}}

\makeatother

\begin{document}

\title{Magnetic Excitations of Undoped Iron Oxypnictides}

\author{Dao-Xin Yao and E. W. Carlson}

\affiliation{Department of Physics, Purdue University, West Lafayette, IN 47907}

\date{\today}

\begin{abstract}
  We study the magnetic excitations of undoped iron oxypnictides using
  a three-dimensional Heisenberg model with single-ion anisotropy.
  Analytic forms of the spin
  wave dispersion, velocities, and structure factor are given. 
  Aside from quantitative comparisons  which can be made to
  inelastic neutron scattering experiments, we also give
  qualitative criteria which can distinguish various
  regimes of coupling strength.    
 The magnetization reduction due to quantum zero point fluctuations shows clear dependence on the c-axis coupling.   
\end{abstract}
\maketitle

The discovery of a new class of superconductors
with transition temperatures exceeding 55K 
has spurred new hope of developing a unified theory of high temperature superconductivity.\cite{kamihara08,ren55k,xu56k}
Like the cuprate superconductors, in the iron pnictide compounds
superconductivity arises from doping a layered antiferromagnet,
giving rise to tantalizing similarities in the phase diagrams. 
However, there are also striking differences.  
For example, whereas importance is placed on 
a single $d$-orbital per Cu site in the cuprates,
in the iron-based superconductors importance
is placed on several $d$-orbitals per Fe site,
and the total nominal spin per site may be large.
Furthermore, the parent compounds of the iron-based
materials are semi-metals, rather than Mott insulators
as in the cuprate case.  
In addition, recent experiments have shown that the
electronic couplings in the iron-based superconductors
are more three-dimensional than in the
cuprate superconductors.\cite{zhao08srfeas,boothroyd08,yuan09}

Although static magnetism tends not to survive in the superconducting
state of the iron pnictides, magnetic excitations have been shown to
play an important role in the superconducting state.  In particular, a
resonance peak has been associated with superconductivity, suggesting
a further connection with cuprate physics.  Because of the
prominent role of magnetism in these materials and the connection of
magnetic fluctuations to the superconducting state, it is important to
understand the simpler magnetic excitations which are present in the
parent compound.  In order to understand the magnetic excitations, we
consider an effective Heisenberg model with exchange couplings between
the net spin associated with each site.  The effective Heisenberg
model may be thought to arise from exchange associated either with
localized magnetic moments, or associated with the net moment arising
from an SDW associated with itinerant electrons.

\begin{figure}[t]
\begin{center}
\resizebox*{0.8\columnwidth}{!}{\includegraphics{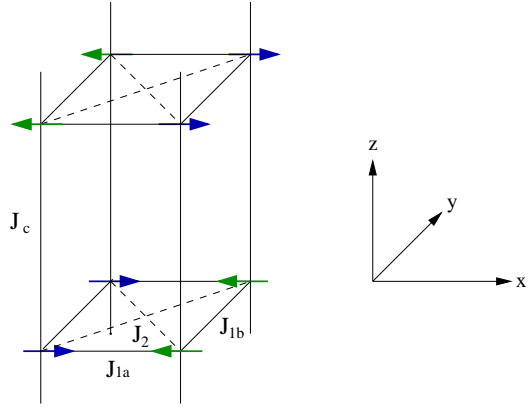}}
\end{center}
\caption{(Color online) Typical magnetic structure of undoped iron-based superconductors.}
\label{lattice}
\end{figure}

At room temperature, most undoped iron-pnictide superconductors have a
tetragonal paramagnetic phase. Upon decreasing temperature, the materials
show a structural transition from tetragonal to orthorhombic. In the 122
materials, a three-dimensional long-range antiferromagnetic order
develops simultaneously. 
This phenomenology constrains effective exchange constants in the Heisenberg model,
written as \cite{zhao08srfeas}
%EC fixed this equation
\begin{eqnarray}
H=&&J_{1a}\sum_{i,j} \mathbf{S}_i\cdot\mathbf{S}_j+J_{1b}\sum_{i,j}
\mathbf{S}_i\cdot\mathbf{S}_j+J_2\sum_{i,j} \mathbf{S}_i\cdot
\mathbf{S}_j  \nonumber   \\
+&&J_c\sum_{i,j}\mathbf{S}_i\cdot\mathbf{S}_j-J_s\sum_{i}(S_i^z)^2 
\end{eqnarray} 
where $J_{1a}$ and $J_{1b}$ are the
nearest neighbor interactions along the a- and b-axes, $J_2$ is the next
nearest neighbor interaction within the plane, $J_c$ is the interaction
along c-axis, $J_s$ is the single ion anisotropy. 
These couplings are illustrated in Fig.~\ref{lattice}.  We use 
linear spin wave theory to study the magnetic excitations and
sublattice magnetization reduction due to quantum zero point
fluctuations. 

%The results are presented
%for the two experimentally argued cases: (1) $J_{1a}=J_{1b}$ case, (2)
%$J_{1a}>>J_{1b}$. We notice the results are very different for the two
%cases.

%%%%%%%%%%% FIGURE %%%%%%%%%%%%%%%%%%%%%%%%%%%%%%%%%%%
%%%%%%%%%%%%%%%%%%%%%%%%%%%%%%%%%%%%%%%%%%%%%%%%%%%%%%

We use Holstein-Primakoff bosons to rewrite the above Hamiltonian as \cite{yaoerica08}
\begin{equation}
  H=E_{\rm Cl}+\sum_{\mathbf{k}} [A_{\mathbf{k}}
  a_{\mathbf{k}}^{\dagger}a_{\mathbf{k}}+\frac{1}{2}(B_{\mathbf{k}}
  a_{\mathbf{k}}^{\dagger}a_{-\mathbf{k}}^{\dagger}+B_{\mathbf{-k}}^*
  a_{\mathbf{k}}a_{-\mathbf{k}})],
\end{equation}
%EC I have changed the sign of all but J_s below.  Please check!
where $E_{\rm Cl}=(-J_{1a}+J_{1b}-2J_2-J_c-J_s)NS^2$ is the classical ground state energy.
The Hamiltonian can be diagonalized by using the Bogoliubov transformation
\begin{equation}
  b_{\mathbf{k}}=\cosh{\theta_{\mathbf{k}}a_{\mathbf{k}}}-\sinh{\theta_{\mathbf{k}}} a_{-\mathbf{k}}^{\dagger},
\end{equation}
which leads to the result 
\begin{equation}
H=E_{\rm Cl}+E_{o}+\sum_{\mathbf{k}}\omega(\mathbf{k})b_{\mathbf{k}}^{\dagger}b_{\mathbf{k}}
\end{equation}
where $\omega(\mathbf{k})$ is the spin wave dispersion and $E_0$ is the quantum zero-point energy correction.

%EC added the factor of "S" in \omega and in E_o.  Please check!  %DY ``S'' has been absorbed into Ak, Bk, no need to put again
The spin wave dispersion $\omega(\mathbf{k})$ is given by
\be
\omega(\mathbf{k})= \sqrt{A_k^2-B_k^2},
\ee
where
\ba
A_k&=&2S (J_{1a} - J_{1b} + 2 J_2 + J_s + J_z + J_{1b} \cos{k_y}), \\
B_k&=&2S (J_{1a} + 2 J_2 \cos{k_y})\cos{k_x} + 2 J_c \cos{k_z}.
\ea
The quantum zero-point energy 
%EC2 correction 
% i.e. it's not a correction -- it just is.
is then
\be
E_0=\frac{1}{2}\sum_{\mathbf{k}}(-A_{\mathbf{k}}+\omega(\mathbf{k})).
\ee

%%%%%%%%%%%% FIGURE %%%%%%%%%%%%%%%%%%%%%%%%%%%%%%%%%%%
\begin{figure}[t]
%\psfrag{kx}{$k_x$}
%\psfrag{ky}{$k_y$}
%\psfrag{w}{$\omega$}
{\centering
  \subfigure
  {\resizebox*{!}{0.7\columnwidth}{\LARGE{(a)}
  \includegraphics{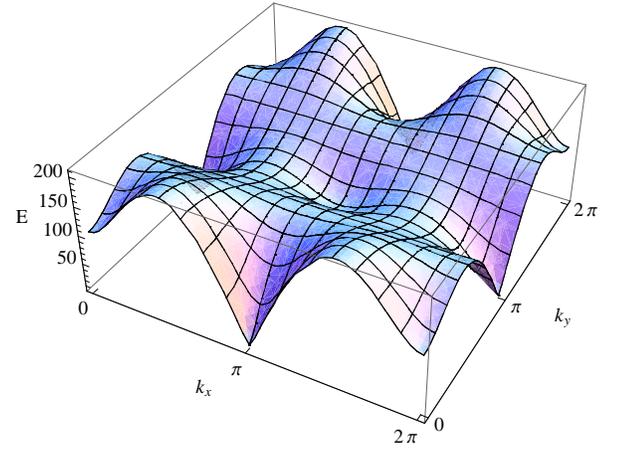}\label{3d1}}}
  \subfigure
  {\resizebox*{!}{0.7\columnwidth} {\LARGE{(b)}
\includegraphics{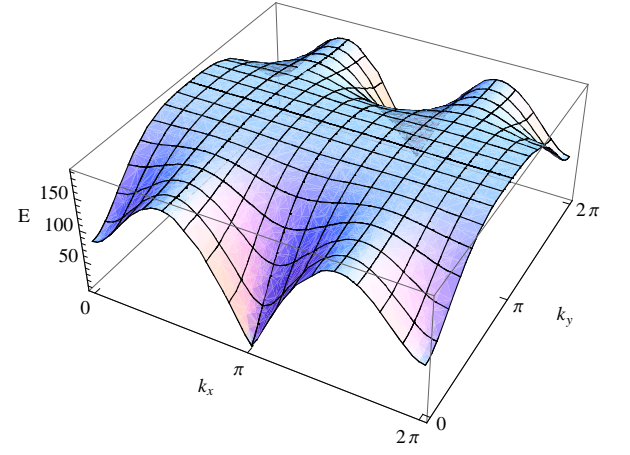}\label{3d2}}} 
  \par}
  \caption{(Color online) Spin-wave dispersion band for the antiferromagnet
shown in Fig.~\ref{lattice}.  (a) Dispersion for $J_{1a} = J_{1b}$, which corresponds to  $J_{1a}=25$, $J_{1b}=25$, $J_2=36$, $J_c=7$, and $J_S=0.05$. 
(b) Dispersion with $J_{1a} \gg J_{1b}$, which corresponds to $J_{1a}=40$, $J_{1b}=-5$, $J_2=20$, $J_c=5$, and $J_S=0.05$.}
\label{spinwave3d}
\end{figure}
%%%%%%%%%%%%%%%%%%%%%%%%%%%%%%%%%%%%%%%%%%%%%%%%%%%%%%%

The  presence or absence of
gaps at particular points in the Brillouin zone may be used to 
gain qualitative information about the state of the system:
\ba \Delta(\pi, 0,
\pi)&=&2S \sqrt{J_s (2 J_{1a} + 4 J_2 + J_s + 2 J_c)},\\  \nonumber
\Delta(0,
\pi, \pi)&=&2S \sqrt{(2 J_{1a} - 2 J_{1b} + J_s) (-2 J_{1b} + 4 J_2 +
J_s + 2 J_c)},\\   \nonumber
\Delta(\pi, \pi, \pi)&=&2S \sqrt{(-2 J_{1b} + 4 J_2 +
J_s) (2 J_{1a} - 2 J_{1b} + J_s + 2 J_c)}, \\   \nonumber
\Delta(0, 0,
\pi)&=&\Delta(\pi, 0, \pi).  \ea 
For example, there can only be a gap at 
$\Delta(\pi, 0,\pi)$ if single-ion anisotropy is present.
Measuring a finite gap at this point requires that $J_s$
be nonzero.  In SrFe$_2$As$_2$, it has been shown that
single-ion anisotropy is present, although it is a very
weak energy scale, $J_s \approx 0.015$meV.\cite{zhao08srfeas}
Given that $J_s$ is a small energy scale, we see
that measuring a gap at $\Delta(0,\pi, \pi)$ would indicate
that there is anisotropy in the electronic degrees of freedom,
{\em i.e.} $J_{1a} \ne J_{1b}$.  In this sense, the value of 
$\Delta(0,\pi, \pi)$ may be taken as a measure of 
electron nematicity in the system.  
It was furthermore established in Ref.~\onlinecite{zhao08srfeas}
that $J_c \gg J_s$.  When this is the case, the magnitude
of the gap at $\Delta(\pi, \pi, \pi)$ may be used to 
indicate proximity to the magnetic phase transition at 
$J_{1b} = 2 J_2$. %DY changed J_1 to J_{1b}  
Because the gaps must be real-valued, %EC2
%EC2 positive or zero, 
we see that the system is constrained to have 
$J_{1b} \le {\rm Min}[J_{1a} + \frac{J_s}{2}, 2 J_2 + \frac{J_s}{2}]$.
Violation of this constraint indicates a change in the
ground state.

In the limit of vanishing single ion anisotropy, 
the gap at  $(\pi,0,\pi)$ disappears, and the following 
spin wave velocities can be defined
\ba
v_x &=& 2 S \sqrt{(J_{1a}+2J_2) (J_{1a}+2 J_2+J_c)}, \\
v_y &=& 2 S \sqrt{(2J_2-J_{1b}) (J_{1a} + 2 J_2 + J_c)}, \\
v_z &=& 2 S \sqrt{J_c (J_{1a}+2 J_2+ J_c)}.
\ea
Notice that $v_y$ becomes imaginary for $J_{1b} > 2J_2$,
indicating a change in the classical ground state configuration.
This is consistent with the stability condition noted above
which is imposed by requiring that $\Delta(0,\pi,\pi)$ remain
real-valued.

First principles calculations of the electronic structure
have made two broad categories of predictions:
(1) $J_2 > J_{1a} \simeq J_{1b}$\cite{yildirim08,ma08},
and  (2) $2 J_2 \simeq J_{1a} \gg J_{1b}$.\cite{yin08,savrasov09} 
Recent neutron scattering experiments have been used to measure the exchange
couplings.  However, their results are quite
different.\cite{mcqueeney09,jun09a,boothroyd08} Here we provide further
predictions to aid in distinguishing the two cases. 
Figures \ref{3d1} and \ref{3d2} show the typical spin wave spectrum
for the two cases. 
In case (1), there are two small spin wave gaps at both $(\pi, 0, \pi)$ and $(0,
\pi, \pi)$. If the system is twinned, two spin gaps may be
observed. However, in case (2) the large interaction anisotropy pushes
the spin wave gap at $(0, \pi,\pi)$ up to the high energy,
which forming a flat zone boundary for case (2), and only one low energy
spin wave gap is expected.

%%%%%%%%%%%% FIGURE %%%%%%%%%%%%%%%%%%%%%%%%%%%%%%%%%%%
\begin{figure}[t]
{\centering
  \subfigure[$J_{1a}=J_{1b}$]{\resizebox*{!}{0.7\columnwidth}{\LARGE{(a)} \includegraphics{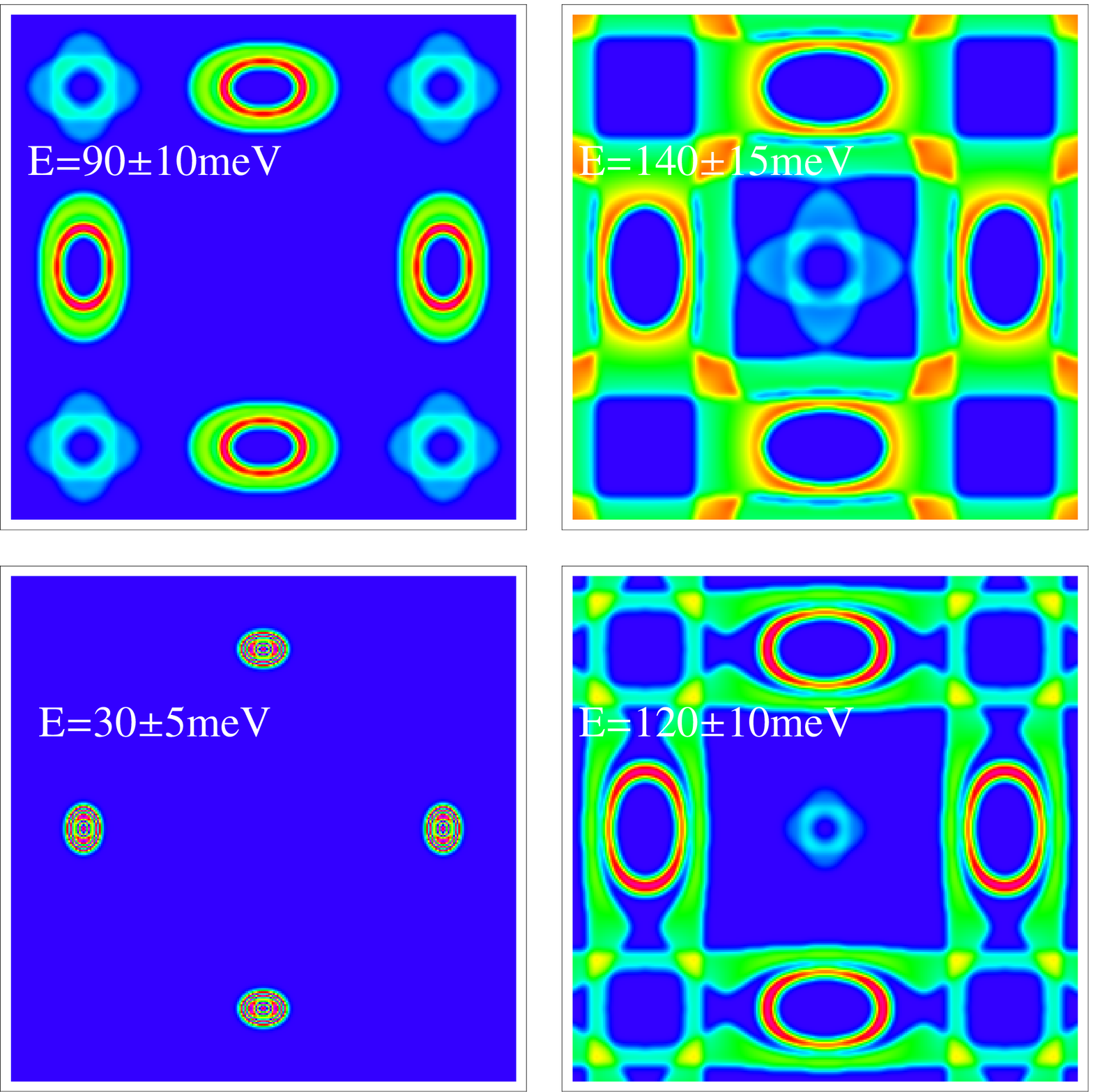}\label{cuts1}}}
  \subfigure[$J_{1a}\gg J_{1b}$]{\resizebox*{!}{0.7\columnwidth}{\LARGE{(b)}\includegraphics{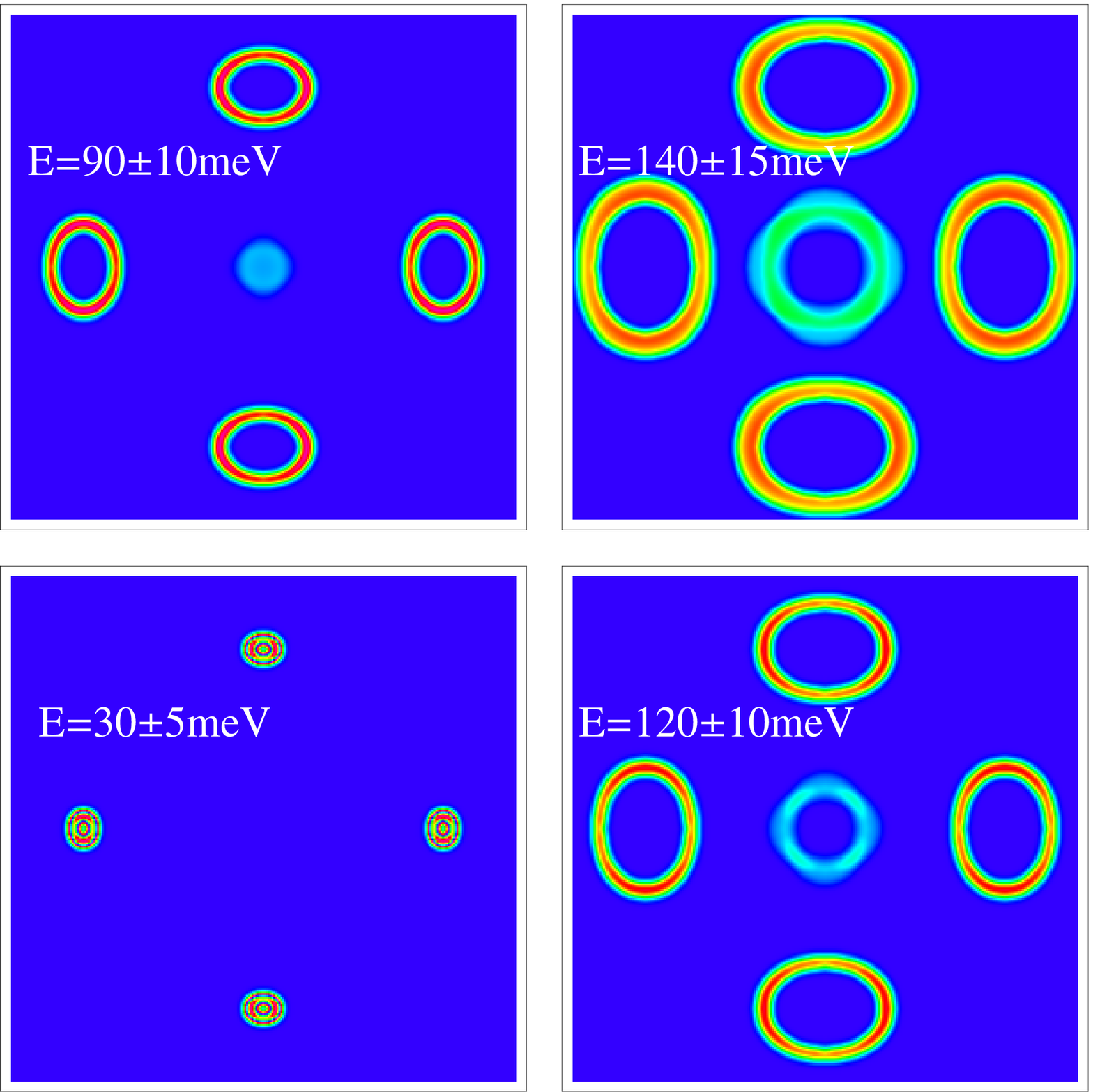}\label{cuts2}}} 
  \par}
\caption{(Color online) Constant-energy cuts (twinned) 
of the dynamic structure factor $S(\mathbf{k},\omega)$ 
for (a) $J_{1a}=J_{1b}$ and  (b) $J_{1a}\gg J_{1b}$. 
The x-axis and y-axis correspond to $k_x$ and $k_y$ respectively
with the range $(-1.5 \pi, 1.5 \pi)$. Interaction parameters are same as Fig.~\ref{spinwave3d}.}
\label{cuts}
\end{figure}
%%%%%%%%%%%%%%%%%%%%%%%%%%%%%%%%%%%%%%%%%%%%%%%%%%%%%%%

The neutron scattering cross section is proportional to the dynamic structure factor $S(\mathbf{k},\omega)$.\cite{boothroyd08} 
In the linear spin-wave approximation, the
transverse parts contribute to the structure factor. 
By symmetry, we have \be S^{xx} (\mathbf{k}, \omega)=S^{yy}
(\mathbf{k}, \omega)=g^2\mu_B^2S_{eff}\frac{A_k-B_k}{2\omega(\mathbf{k})}
[n(\omega)+1]\delta(\omega-\omega(\mathbf{k})), \ee where $S_{eff}$ is the
effective spin on an Fe ion, $g$ is the g-factor of iron ($\sim 2$), and  $n(\omega)$ is the Bose occupation factor.
% \be
%\frac{d^2\sigma}{d\Omega dE_f}=\frac{k_f}{4k_i} (\gamma r_0)^2 g^2 f^2
%(k)\exp{(-2W)} \sum_{\alpha
%\beta}(\delta_{\alpha,\beta}-\hat{k}_{\alpha}\hat{k}_{\beta})
%S^{\alpha \beta} (\mathbf{k}, \omega),
%\ee 
%where $k_i$ ($k_f$) is the
%moment of incident (outgoing) neutron, $g$ is the g-factor of iron,
%$\exp{(-2W)}$ is the Debye-Waller factor (close to ``1'' at low
%temperature), and $S^{\alpha \beta} (\mathbf{k}, \omega)$ is the
%structure factor. In the linear spin-wave approximation, the
%transverse part contribute to the structure factor. Based on the
%rotational symmetry, we have \be S^{xx} (\mathbf{k}, \omega)=S^{yy}
%(\mathbf{k}, \omega)=S_{Fe}\frac{A_k-B_k}{2\omega(\mathbf{k})}
%[n(\omega)+1]\delta(\omega-\omega(\mathbf{k})), \ee where $S_{Fe}$ is the
%effective spin of Fe ion (equals "S'' in the linear spin wave theory),
%$n(\omega)$ is the bose factor.

In Fig.~\ref{cuts} we show intensity plots at constant energy for the
dynamic structure factor $S(\mathbf{k},\omega)$, assuming a crystal
with twinned antiferromagnetic domains.  In the presence of twinning,
two concentric %EC3 Daoxin, do you agree with the word "concentric"?
spin wave rings are expected at low energy if the neutron scattering
resolution is high enough for case (1). At high energy, the outer ring
increases quicky for case (1) and can form bright spots as the rings
merge.  In case (2), only one ring will be observable since the energy
gap at $(0, \pi, \pi)$ goes to very high energy.  In addition, the
band top in case (2) becomes flat in a very large portion of the
Brillouin zone.  (See Fig.~\ref{3d2}).

%%%%%%%%%%%% FIGURE %%%%%%%%%%%%%%%%%%%%%%%%%%%%%%%%%%%
\begin{figure}[t]
\begin{center}
\resizebox*{0.8\columnwidth}{!}{\includegraphics{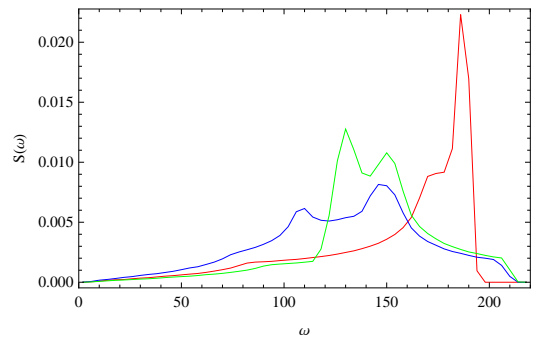}}
\end{center}
\caption{(Color online) S($\omega$) for the full Brillouin zone. The blue line
corresponds to case (1) $J_{1a}=J_{1b}$, the red line corresponds to
$J_{1a}\gg J_{1b}$, and the green line is for an intermediate case
which has $J_{1a}=40$, $J_{1b}=20$, $J_2=30$, $J_c=5$, $J_s=0.05$.
Interaction parameters in cases (1) and (2) are same as Fig.~\ref{spinwave3d}. $S(\omega)$ is in units of $g^2\mu_B^2 S_{eff}$.} %DY a new intermediate
%anisotropic case added
\label{sw}
\end{figure}
%%%%%%%%%%%%%%%%%%%%%%%%%%%%%%%%%%%%%%%%%%%%%%%%%%%%%%%

The integrated structure factor $S(\omega)$ 
can also be used to distinguish the two cases:
\be
S{(\omega)}^{\alpha \alpha}= \int\int\int_{BZ}dk_xdk_ydk_z S^{\alpha \alpha}(\mathbf{k},\omega) \delta(\omega-\omega(\mathbf{k})),
\ee
where $\alpha=x, y$ and BZ means integrate over the full 
magnetic %EC3 Daoxin, do you agree with the word "magnetic" here?
Brillouin zone. Numerical results are presented in Fig. \ref{sw}.
The most dramatic differences are expected in the high energy response.
There are two broad peaks expected for the $J_{1a}= J_{1b}$
case. There is a sharp peak at high energy for the $J_{1a}\gg J_{1b}$ case
which is caused by the very large density of states near the spin wave
zone boundary. As we see from Fig.~\ref{3d2}, the spin wave band is flat in a large portion of the Brillouin zone.  We also show a curve of $S(\omega)$ for the case of $J_{1a}=2J_{1b}$ ($J_{1a}$ is not much bigger than $J_{1b}$).

The total moment sum rule for a Heisenberg model with spin S
is defined as \cite{lorenzana05} 
\ba M_0&=&\frac{1}{N}\sum_{\alpha}
\int d\mathbf{k}\int_{-\infty}^{\infty} d \omega S^{\alpha \alpha}
(\mathbf{k},\omega) \nonumber \\
 &=&M^x+M^y+M^z \nonumber \\
&=&g^2\mu_B^2 S(S+1).  \ea 
The transverse part of this corresponds to 
$M^{\alpha}= \frac{1}{N}\int\int\int_{BZ} dk_xdk_ydk_z \int
d\omega S^{\alpha \alpha}(\mathbf{k},\omega)$, where $\alpha=x, y$. We
get $M^x=M^y=0.61 g^2 \mu_B^2 S_{eff} $ per Fe for case (1) and $M^x=M^y=0.58
g^2 \mu_B^2 S_{eff}$ per Fe for case (2). 

%EC3
%The elastic dynamic spectral weight
%is $2.56 \mu_B^2$ %EC3 -- it was .64, but we needed the g^2.
%per Fe from elastic neutron scattering.\cite{somebody}
%Thus the total moment $M_0$ is $1.72 \mu_B^2$ for the
%case (1) and $1.66 \mu_B^2$ for case (2).
%*** {\bf We must have $S_{\rm Fe}$ in order to calculate this -- what are we assuming?}
%Also, the sum rule only applies if there's no double occupancy, 
%and the sum rule does not include the polarization factors.  
%***

%EC3 These two predictions will be indistinguishable experimentally!  Should we say that it is 
%a way to distinguish local moment from itinerant electrons?

\begin{figure}[t]
\begin{center}
\resizebox*{0.8\columnwidth}{!}{\includegraphics{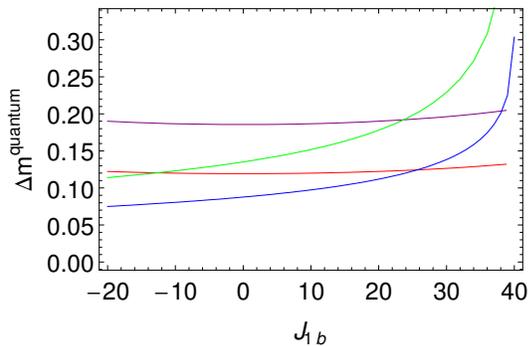}}
\end{center}
\caption{(Color online ) $J_{1b}/J_{1a}$ dependence of the reduction of the sublattice
 magnetization due to zero point energy of the spin waves. Red
 ($J_c=5$ meV) and purple ($J_c=0$ meV) lines corresponds to
 $J_{1a}=J_{1b}$, blue ($J_c=5$ meV) and green ($J_c=0$) lines
 correspond to $J_{1a}\gg J_{1b}$.  Interaction parameters are same as Fig.~\ref{spinwave3d}.}
\label{dmag}
\end{figure}

The measured magnetic moment per iron is typically less than
one Bohr magneton ($\mu_B$), which is much smaller than the theoretically expected
value of $\sim2.3 \mu_B$ per iron site predicted by LDA calculations.\cite{cao08,ma08,dongwang08}
In spin-wave theory, both the quantum zero point fluctuations and thermal fluctuations
reduce the expected magnetic moment per site. Since the energy scale of
iron-based supersonductors is much larger than the temperature scale,
we focus on the the quantum zero point fluctuations. The sublattice
magnetization reduction $\Delta_m^{quantum}$ is defined as
\begin{equation}
  \Delta m^{quantum} =\frac{1}{2}\int_0^{2\pi} \int_0^{2\pi}\int_0^{2\pi}
  \frac{dk_x}{2\pi} \frac{dk_y}{2\pi}
  \frac{dk_z}{2\pi}\frac{A_{\mathbf{k}}}{\omega(\mathbf{k})}
  -\frac{1}{2}.
\end{equation}
In Fig. \ref{dmag} we  present the numerical results for both cases (1) and (2).  
Note that $J_c$ has an important effect on the magnetization reduction.
Empirically, the 122 materials are
generally more three-dimensional than the 1111 materials.\cite{lynn09}
Thus the effective magnetic
moment for 122 material is expected to be higher than 1111
materials. If $J_c\approx 5meV$, we have $\Delta m^{quantum}\approx
0.1$. For $S=1/2$, this gives $S_{eff}\approx 0.4$, which is close to
the magnetic moment found by experiments in the 122 materials, which
have $g S_{eff}\approx 0.8 \mu_B$.~\cite{lynn09} %DY

%%%%%%%%%%%% FIGURE %%%%%%%%%%%%%%%%%%%%%%%%%%%%%%%%%%%
%%%%%%%%%%%%%%%%%%%%%%%%%%%%%%%%%%%%%%%%%%%%%%%%%%%%%%%

 In summary, we have used the three-dimensional Heisenberg model and spin
  wave theory to study the magnetic excitations, dynamic structure
  factor, and sublattice magnetization for the antiferromagnetic spin
  state found in the undoped iron-based superconductors.  The anisotropy of exchange couplings
  within the Fe-As plane can cause clear differences in the spin
  wave spectrum, expected scattering intensity at constant energy, and integrated dynamic
  structure factor. These can be used to 
  determine the degree of anisotropy of magnetic exchange interactions
  within the Fe-As plane.  In addition, we have calculated the sublattice magnetization reduction
  from the quantum zero point fluctuation. The results show that $J_c$ can enhance the long range magnetic ordering
  dramatically, which is consistent with the experimental findings about the magnetic moments in 1111 and 122 materials.

\begin{acknowledgments}
We thank J. P. Hu, P. Dai, J. Zhao, S. Li and E. Dagotto for helpful discussions.  This
work was supported by Research Corporation and by NSF Grant No. DMR
08-04748.

\end{acknowledgments}

\end{document}